\documentclass[%
 aip,
jcp,%
preprint,%
]{revtex4-1}

\usepackage{graphicx}
\usepackage{dcolumn}
\usepackage{bm}
\usepackage[percent]{overpic}

\begin{document}

\title{Band bending at the interface in Polyethylene-MgO nanocomposite dielectric}%

\author{Elena Kubyshkina}
\email{elena.kubyshkina@ee.kth.se}
\affiliation{KTH Royal Institute of Technology, School of Electrical
Engineering, Electromagnetic Engineering, SE-10044 Stockholm, Sweden
}
\author{Mikael Unge}
\affiliation{ABB Corporate Research, SE 72178 V\"aster\r{a}s, Sweden}
\altaffiliation{KTH Royal Institute of Technology, School of Chemical Science and Engineering, Fibre and Polymer Technology, Stockholm, SE-100 44 Sweden}
\author{B. L. G. Jonsson}
\affiliation{KTH Royal Institute of Technology, School of Electrical
Engineering, Electromagnetic Engineering, SE-10044 Stockholm, Sweden
}

%

\date{\today}

\begin{abstract}
Polymer nanocomposite dielectrics are promising materials for electrical insulation in high voltage applications. However, the physics behind their performance is not yet fully understood. We use density functional theory to investigate electronic properties of the interfacial area in magnesium oxide-polyethylene nanocomposite.  Our results demonstrate polyethylene conduction band matching with conduction bands of different surfaces of magnesium oxide. Such band bending results in long range potential wells of up to 2.6 eV deep. Furthermore, the fundamental influence of silicon treatment on magnesium oxide surface properties is assessed. We report a reduction of the surface-induced states at the silicon-treated interface. The simulations provide information used to propose a new model for charge trapping in nanocomposite dielectrics.
\end{abstract}

\maketitle
Understanding the physics at interfaces between nanoparticles and bulk matrix in polymeric nanocomposite dielectric materials will bring us closer to the possibility of tailoring high voltage insulation with desired properties. Elucidation of the charge trapping phenomena in nanocomposite dielectrics is a
 step towards comprehending the complex electrical behavior of these
materials. Charge carriers are inevitably present in dielectrics due to
impurities and charge injections from electrodes. Mobile electrons can  cause
polymer bond breakage or impact ionization and eventually lead to degradation of
the material \cite{Sun2014}. Deep trapping sites reduce electron mobility and therefore improve the performance of insulators \cite{Teyssedre2005, Unge2015}. However, deep traps also enhance charge injection which could lead to a higher conductivity \cite{Huzayyin2011}. Thus, trapping phenomenon are of significant importance for electrical properties of dielectrics.

There are several theories explaining the origins of traps in nanocomposites.
One of the models \cite{Takada2008} proposes that electrons are trapped in electric potential wells caused by induced dipoles of spherical nanoparticles.
A well-known multi-core model that was proposed by Tanaka \cite{Tanaka2005} suggests three layers of polymer with altered properties that surround a nanoparticle. According to the model, these layers possess a number of shallow and deep traps due to a ``less ordered'' polymer structure. Additionally, polymer morphology was demonstrated \cite{Unge2012} to be important for polyethylene band gap value.
Furthermore, interfaces may exhibit under-coordinated atoms, which result in trap states \cite{Kittel,Callaway}. These states are highly dependent on the chemistry of nanoparticle surface modification \cite{Dongling2005}. Surface modification is silicon based; it creates a stable bonding between nanoparticles and polymer and contributes to better particle dispersion in the matrix. Silicon-treated nanoparticles are less hydrophilic and provide the nanocomposite with a higher resistivity in a humid environment \cite{Pallon2016}. Ab-initio modelling indicated \cite{Shi2008} defect states and band bending at the vinlysilanediol-treated SiO$_2$-polyvinylidene difluoride interface.

We focus here on polyethylene -- magnesium oxide nanocomposite. This choice is due to the fact that it has been experimentally shown to have better dielectric behavior in high-voltage direct current applications as compared to pure polymer material \cite{Pallon2016, Zhang2014, Murakami2008}. These experiments show that the addition of MgO nanoparticles suppresses space charge formation in the material and that the nanocomposite has a higher partial discharge resistance, increased relative permittivity and decreased conductivity. Despite the fact that bulk magnesium oxide has a relatively simple rock salt cubic crystal structure, its surface is complex and irregular. The chemical reactivity varies with surface structure such as steps and edges \cite{Sushko2002}. Surface defects bring additional energy levels to the band gap. For example, oxygen vacancies with one and two electrons in the cavity of the vacancy, introduce traps up to 2.87 eV and 3.38 eV respectively \cite{Florez2008}. These defects can be formed during dehydroxylation at high temperatures \cite{Coluccia1988} –- a common procedure in the preparation of nanoparticles used for nanocomposite test samples. In addition to that, the manufactured particles are polycrystalline. The grain boundaries in polycrystalline materials have been shown to trap electrons \cite{Seto1975}. For MgO in particular such traps were calculated to be 1 eV deep \cite{McKenna2008}.

To investigate the electronic properties of the interfacial region in the nanocomposite, we consider interfaces between perfect defect-free MgO surfaces and polyethylene. We limit our study to two surface planes: MgO (100) and hydroxylated MgO (111). These surfaces were shown \cite{Wander2003, Refson1995} to be the most stable under dry and aqueous conditions respectively. The high stability of these surfaces ensures their wide presence in the material. To assess the effect of silicon treatment, we have also have studied interface with silicon-bonded polyethylene to MgO (111) surface. In this letter we ignore the polymer morphology, and limit our study to the effects of the contact between a crystalline surface and a polymer-like molecule.

We exploit density functional theory as implemented in Vienna Ab initio Simulation
Package (VASP) \cite{Kresse1993, *Kresse1994, *Kresse1996, *Kresse1996_b}. The hybrid functional HSE06 \cite{Krukau2006} was used for this study.
To investigate the role of Van der Waals correction several trial calculations were performed with the PBE functional \cite{Perdew1996, *Perdew1997}. The Tkatchenko-Scheffler correction \cite{Tkatchenko2009} did not provide any significant ($\leq$0.1 eV) change in the densities of states (DOS) and geometries of the interfaces calculated with PBE. Therefore van der Waals correction was not used for the current study.
As we model non-symmetrical slabs, we used dipole correction as implemented in VASP \cite{Makov, Neugebauer}. K-mesh was set to 3x3x1, and cutoff energy to 400 eV. 

\begin{figure}
\begin{overpic}[scale=1.52]{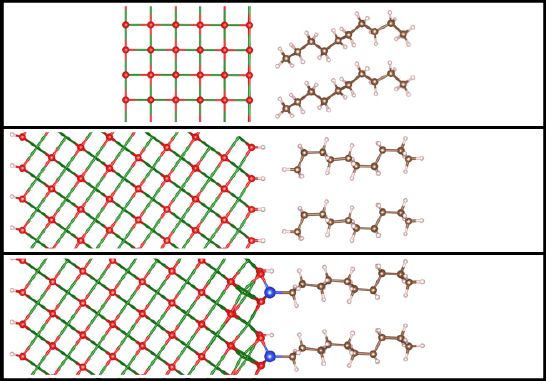}
\put (94,65) {a.}
\put (94,42) {b.}
\put (94,19) {c.}
\end{overpic}
\caption{Geometries representing the modeled systems:
 a) \label{geom} MgO (100)--D-100 interface;
 b) h-MgO (111)--D-111 interface;
 c) MgO (111)--Si--D-111  interface.
 Magnesium atoms are shown in green, oxygen in red, hydrogen in white, carbon in brown, silicon in blue colors \cite{VESTA}.}
\end{figure}

We consider three interfacial configurations, with geometries as depicted in Figure \ref{geom}.
Polyethylene -- MgO (100), the unit cell consists of six layers of MgO and one C$_{10}$H$_{22}$ chain as polyethylene model.
Polyethylene -- hydroxylated MgO (111), the unit cell consists of nine layers of MgO and a C$_{10}$H$_{22}$ chain corresponding to polyethylene. The bottom of the crystal slab was saturated with hydrogen to comply with the requirement of charge neutrality.
Polyethylene -- Silicon-treated MgO (111), the system is represented by the following structure: three out of four oxygen surface atoms are bonded through a Si atom to the alkane chain C$_{10}$H$_{21}$, and the remaining oxygen is saturated with hydrogen.
The thickness of the MgO slabs was chosen such that at least two  layers in the middle exhibit the same density of states including core levels. Three of the bottom layers of MgO were fixed during the structure relaxation to mimic bulk behavior. 

As intermediate and reference models the following structures were considered: two different configurations of aligned C$_{10}$H$_{22}$ chains in vacuum representing polyethylene for MgO (100) and MgO (111) cases, MgO (100) surface, hydroxylated MgO (111) surface and --SiOH-terminated MgO (111) surface. All of the models were supplemented with a reasonable vacuum region. All the calculated data below are presented for the relaxed systems, i.e. for the final structures corresponding to the energy minima. The structures had been relaxed until atomic forces were below 0.05 eV/\AA.

The most straightforward way to assess the relative positions and shifts of the polyethylene and MgO at the interface is to align them to the vacuum level. However, the position of the vacuum level is highly dependent on the surface induced electronic polarization \cite{Logsdail2014, Logsdail2015} and the vacuum level position explicitly extracted from the calculations is not reliable for the structures with MgO in the unit cell. Also, the vacuum level is mostly influenced by the atoms closest to the vacuum region. In our geometry the interface is too ``far'' from the vacuum to influence its energy. Therefore we used the following procedure to locate the positions of the bands with respect to the vacuum level. First, we found the position of the valence band of the bulk-like region of MgO (100) slab (-5.7 eV), then  corrected it with multipolar shift $D_s$ calculated in Ref.~\citenum{Logsdail2014}, averaged to 0.9 eV ($\pm$ 0.1 eV). We fixed this value for all the systems and used it as a reference. 

The band energies with respect to the valence band of bulk MgO are presented in Table \ref{tab:table1}. We define the band edges as the values at which the curve of density of states reaches zero. Such approach results in possible errors of about 0.1 eV in conduction and valence band energies due to the limited precision of density of states.
\begin{table}
\caption{\label{tab:table1}Calculated band gaps,  positions of valence and conduction bands (VB and CB) of the studied systems, with respect to the vacuum level, eV.}
\begin{tabular}{lccc}
    & VB position  & CB position  & Band Gap \\
\hline
PE&&& \\ \hline
PE for (100)&-5.9&0.5&6.3\\
interface  (100)   &  -8.0  & -1.7  &  6.3\\
PE for  (111) &-7.4&-0.9&6.5\\
interface  (111) Si     &  -7.9  & -1.4  &  6.5 \\
interface (111) OH   &  -10.0 &  -3.5  &  6.5\\ \hline
MgO&&& \\ \hline
bulk-like MgO  &-6.6 & -0.7 & 5.9 \\
surface  (100)  &  -6.6  &-1.8   &  4.8\\
interface  (100)   &  -6.6 &  -1.8  &  4.8\\
surface  (111) OH  &  -6.8 &  -3.6  &  3.2\\
interface  (111) OH  &  -6.8  &  -3.5 &  3.3\\
surface  (111) SiOH   &  -6.7  &  -1.6 &  5.1\\
interface  (111) Si       &  -6.8  & -1.4  &  5.4\\

\end{tabular}
\end{table}
First we evaluate the presence of any specific interfacial states due to the contact between magnesium oxide and polyethylene. Here and further on we refer to  C$_{10}$H$_{22}$ as ``polyethylene'' despite of it being a simplified version of the polymer.

 First MgO (100) surface is considered. Our calculations show that it has a 1.1 eV narrower band gap than the bulk oxide (Table \ref{tab:table1}). When polyethylene is introduced to the system, the surface-associated states do not change and polyethylene band gap value is also the same as in the vacuum. However, the position of the valence band with respect to vacuum level is 2.1 eV lower than it is expected from calculations of polyethylene only (``PE for (100)'' in Table \ref{tab:table1}).

\begin{figure}
\includegraphics[trim={2.5cm 3.5cm 2.6cm 0},clip,scale=0.52]{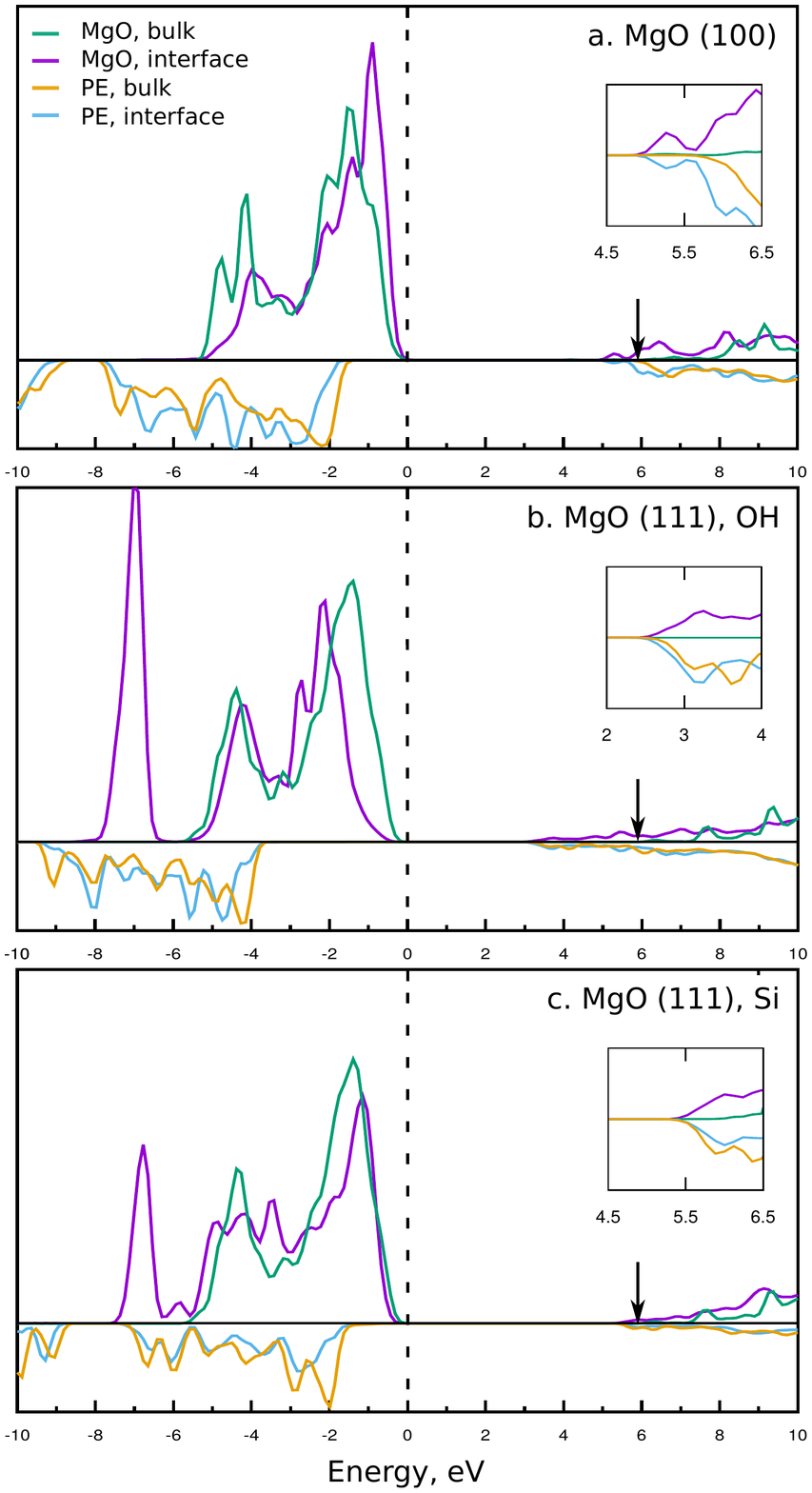}
\caption{\label{dos} Projected densities of states of the interfaces:
 a) MgO (100)--polyethylene interface;
 b) hydroxylated MgO (111)--polyethylene interface;
 c) Si-treated MgO (111)--polyethylene interface.
 States from the atoms in bulk-like region of the MgO slab are shown in green, states from the two upper layers of MgO at the surface + hydrogen in b) and silicon in c) -- in purple, states from atoms in the middle region of polyethylene -- in yellow and states from the two CH$_{2}$ radicals closest to the interface -- in blue color.
 Zero corresponds to bulk MgO valence band (-6.6 eV with respect to the vacuum level). Magnified conduction band edges are shown in squares. Arrows show the bottom of bulk MgO conduction band.
}
\end{figure}

For the polar MgO (111) surface two cases were considered: ``untreated'' hydroxylated surface and --SiOH-saturated surface.
Hydroxylated MgO (111) possesses continuous surface states spread up to 2.7 eV below the conduction band of bulk MgO (Figure \ref{dos}-c). Further analysis of the projected DOS shows that these states are of hydroxyl origin. These results are consistent with previous works \cite{Lazarov2005}. Addition of the polyethylene leads to an insignificant expansion of the band gap by 0.1 eV which is in the order of magnitude of a possible error. The polyethylene bands are 2.6 eV lower than in  vacuum.

 --SiOH-saturated MgO (111) structure is an intermediate step for the polyethylene -- silicon-treated interface. It also provides more understanding to the direct impact of the silicon atom on the electronic structure of the MgO (111) surface. To model this system 3 out of 4 top oxygen atoms were bonded to --SiOH radical and relaxed. The resulting structure has a band gap of 5.1 eV, which is 0.7 eV smaller than the bulk and 1.9 eV larger than the hydroxylated surface.  When the Si-bonded --OH group is replaced by the --C$_{12}$H$_{25}$ chain, we see that the band gap increases by additional 0.3 eV. Thus, silicon treatment removes hydroxyl-associated states in the conduction region, as was also predicted in our earlier works \cite{Kubyshkina2015,Kubyshkina2016}. In this case the polyethylene bands are also shifted, by 0.5 eV.

Furthermore, densities of states of the interface systems (Fig. \ref{dos})
show a remarkable consistent pattern. In all of the cases conduction band of polyethylene and magnesium oxide surfaces are aligned. The position of polyethylene conduction band is solely defined by the position of the conduction band of the surface in the neighbourhood. A possible reason for this behavior is the character of the conduction states in polyethylene and MgO. The polyethylene conduction zone is of interchain character \cite{Serra2000}, i.e. the conduction states are located in vacuum surrounding the chains. In that sense  the studied surfaces are somewhat similar. MgO (100) states slowly diffuse into the vacuum from the surface, and hydroxylated MgO (111) surface also has conduction states outside the slab. As polyethylene and MgO surface states  overlap locally, the interaction among them leads to the matching of their energies.

\begin{figure}
\includegraphics[width=\linewidth]{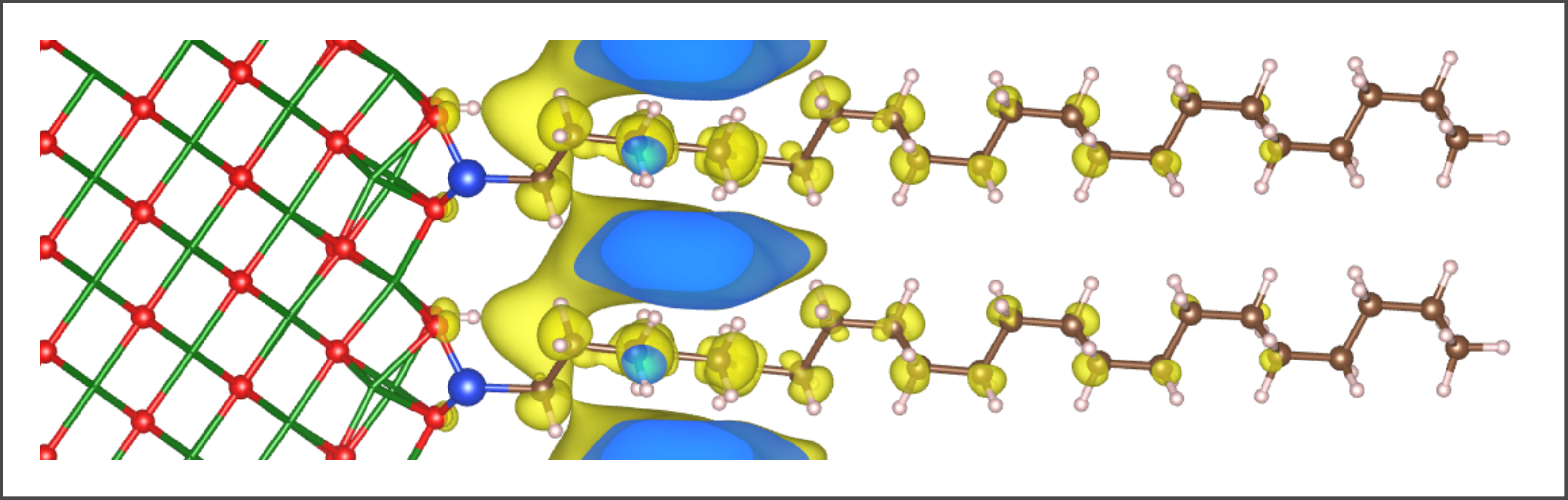}
\caption{\label{states} States in 0.2 eV interval from  the bottom of the conduction band for extended Silicon-treated MgO (111)--polyethylene interface are depicted as yellow isosurfaces with cyan sections.
 Magnesium atoms are shown in green, oxygen in red, hydrogen in white, carbon in brown, silicon in blue colors.
}
\end{figure}

To assess spatial scales of the polyethylene band bending, we considered silicon-treated interface with extended length of polyethylene chain (of 20 --CH$_3$ units). Due to high cost of hybrid functional calculations, this system was calculated with the LDA functional. Our test calculations with the smaller systems has shown that the qualitative behaviour captured with LDA functional\cite{Martin2004} is similar to the HSE06 predictions. The states in 0.2 eV interval from  the bottom of the conduction band are visualized in Fig. \ref{states}.  The projected DOS (not shown) shows 0.1 eV shift of the conduction band  towards its bulk value within  about 15 \r{A} distance from the interface. Assuming linear behavior, e.g. 0.1 eV shifts for every 15 \r{A},  polyethylene conduction band position should return to its bulk value within 45 \r{A}  distance from the Silicon-bonded interface. Assuming the same behavior for the other interfaces, one can predict an upper limit of the band bending spatial scale to be about 40 nm. This is of course an approximation, but it provides an idea of the extent of the nanoparticle influence. Long range potential well is several eV deep with minima at the interface and it will facilitate the movement of an injected electron towards the surface of the particle and its localization there. In the case of an untreated nanoparticle however the surface is a large area of low energy states. Despite of electron being localized with respect to the bulk materials, it won't be localized on the interface. Silicon treatment solves this problem. Silicon bonds will remove some of the hydroxyl-associated states, and the remaining ``untreated'' parts of the surface will act as localized traps. The long range energy well provided with polyethylene at the interface could also explain the sensitivity of the nanocomposite performance to the particle dispersion \cite{Ju2014}. When the dispersion is poor, overlapping interfacial regions create local percolation paths for conducting electrons.

Moreover, the valence band of MgO (100) is slightly higher than that of the bulk, leading to a possible electron deficiency at the surface. In this case, as the electron and hole will be spatially close, recombination becomes a possible scenario. MgO (100) traps are just 0.1 eV deeper than traps associated with grain boundaries. It is then probable for an electron to penetrate inside the particle and possibly recombine there.
Density of states at the hydroxylated MgO (111) surface provides conditions for migration of holes towards the inner part of the particle, as the valence band of the bulk crystal is higher than at the surface. Combined with the conduction band energy, a potential for formation of dipoles is created.

Given these outcomes, we can make three important conclusions. First, the contact between the polymer and the MgO surface does not introduce any new polymer conformation-related states. From Figure \ref{dos} one can see that the lowest energy states in the conduction band originate from the ideal crystal surfaces.
Second, we found that silicon treatment of MgO surface does not only improve particle dispersion and  controls hydrophilicity, but it also drastically changes the electronic structure of the interface.
And third, our calculations demonstrate physical foundations for electrical double layer formation at hydroxylated MgO (111) surface and recombination of the positive and negative charges at MgO (100) surface.
Complex nonlinear behavior of nanocomposites in an electric field could thus be explained by the influence of the double layer potential, as was suggested in Ref. ~\citenum{Lewis2004}.

In conclusion, we propose a new theory for charge trapping in the nanocomposite.
The nanoparticle surface introduces a complex potential with different energies of free-electron states. Whereas interfaces with non-bonded polymer chains provide deep energy surface states, the interface with silicon bonding between the crystal and the polymer has a clear band gap as wide as that of bulk MgO. Polyethylene bands bending forces the excess electrons to the vicinity of the nanoparticle.  These electrons are captured in the surface-provided low energy states. As the states are separated from each other by  regions with silicon-bonded polymer, electron mobility along the surface will also be restricted.  We believe that this model can be extended for other  nanocomposites, especially polyethylene with nanoparticles of negative electron affinity, such as silicon oxide and boron nitride. However further work is needed to verify it.

\begin{acknowledgments}
E. K. gratefully acknowledges discussions with A. Logsdail, A. Sokol and A. Huzayyin and the group of I. Abrikosov for VASP computation tutoring. This work is financially supported by Elforsk, project 36151 and by SweGRIDS. M. U. gratefully acknowledges financial contribution from Swedish Governmental Agency for Innovation Systems (Vinnova), project 2015-06557. The computations were performed on resources provided by the Swedish National Infrastructure for Computing (SNIC) at National Supercomputer Centre.
\end{acknowledgments}

\bibliography{bibliography}
\end{document}